\documentclass[
reprint,
rsi,
amsmath,
amssymb]{revtex4-2}
\usepackage{graphicx}    
\usepackage{dcolumn}     
\usepackage{bm}          
\usepackage{physics}     
\usepackage{xcolor}
\usepackage{siunitx}     
\usepackage{float}       
\usepackage{booktabs}    
\begin{document}

\title{ A Top-Loading Point-Contact Spectroscopy Probe with In-Situ Sample Exchange for Dilution Refrigerators}

\author{Ghulam Mohmad, Atanu Mishra, and Goutam Sheet}

\email{goutam@iisermohali.ac.in}
\affiliation{Department of Physical Sciences, Indian Institute of Science Education and Research (IISER) Mohali, Sector 81, S. A. S. Nagar, Manauli, P.O. 140306, India}
\begin{abstract}
We report the design and implementation of a point-contact spectroscopy (PCS) system integrated with a dilution refrigerator, enabling measurements down to 30 mK. The setup employs a needle-anvil geometry with a cryogenic piezo-driven nanopositioner for in-situ formation of mesoscopic point contacts. We discuss the thermal anchoring strategies that enable efficient cooling of the probe to ultra-low temperatures and reliable measurements. We also address positioner-related challenges and the solutions implemented to ensure stable operation at millikelvin temperatures. The performance of the probe is demonstrated through point contact spectroscopy  on Ta-doped TiSe$_2$ (Ta$_x$Ti$_{1-x}$Se$_2$, $x = 0.2$), a superconductor with $T_c \approx 2.3$ K. The spectra exhibit well-defined superconducting features that systematically diminish with increasing temperature and magnetic field. The platform provides a robust and versatile tool for spectroscopic investigations of superconductors and other quantum materials at millikelvin temperatures and high magnetic fields.
\end{abstract}

\maketitle

\section{Introduction}
Point contact spectroscopy (PCS) is a transport spectroscopic technique that probes energy dependent quasiparticle scattering within a confined constriction between two conductors. The technique relies on the semiclassical description of the flow of charges through a small orifice connecting two bulk conductors. When a bias voltage $V$ is applied across the contact, charges injected from one electrode acquire excess energy $eV$. This provides energy-resolved information. 

For a circular constriction of radius $a$, the nature of transport depends on the ratio $a/l$, where $l$ is the electronic mean free path. In the ballistic limit $a \ll l$, electrons traverse the contact without momentum relaxation. The resistance is then given by the Sharvin expression \cite{popov1963possible, wexler1966size}.

\begin{equation}
R_S = \frac{2h}{e^2} \frac{1}{(k_F a)^2},
\end{equation}

\noindent which depends only on the Fermi wave vector $k_F$ and the contact geometry. Under bias, the electron distribution inside the constriction becomes anisotropic and nonequilibrium. The current is carried by electrons occupying a restricted solid angle in momentum space defined by the contact aperture.

If inelastic scattering processes are present, electrons crossing the constriction may lose energy by exciting elementary modes of the system. The probability of backscattering then becomes energy dependent. Within the semiclassical framework developed by Yanson \cite{yanson1974nonlinear}, the nonlinear correction to the ballistic current is directly related to the energy dependent inelastic scattering rate. One obtains

\begin{equation}
\frac{d^2 I}{dV^2} \propto 
\left. \frac{d}{dE} \left( \frac{1}{\tau(E)} \right) \right|_{E=eV},
\end{equation}

where $\tau(E)$ is the inelastic relaxation time. For electron--phonon interaction,

\begin{equation}
\frac{1}{\tau(E)} \propto \int_0^{E} \alpha^2 F(\omega)\, d\omega,
\end{equation}

so that

\begin{equation}
\frac{d^2 I}{dV^2} \propto \alpha^2 F(eV).
\end{equation}

Here $\alpha^2 F(\omega)$ is the Eliashberg electron--phonon spectral function \cite{duif1989point,daghero2010probing}. Peaks in $d^2I/dV^2$ therefore occur at voltages corresponding to phonon energies $\hbar \omega = eV$. The voltage axis directly maps onto excitation energy. This establishes PCS as a true quasiparticle scattering spectroscopy.

The same principle applies to other inelastic channels. In magnetic materials, scattering from magnons produces conductance features at characteristic magnon energies\cite{naidyuk2019point}. In Kondo systems, spin fluctuation scattering modifies the low bias conductance\cite{jansen1981application}. In rare earth compounds, crystal field excitations give rise to identifiable spectral structures. In each case, PCS measures how the energy dependent scattering rate modifies transport through the constriction.

When $a \sim l$, transport enters the diffusive regime. Both ballistic and resistive contributions appear, and the resistance follows the Wexler interpolation formula ~\cite{wexler1966size}

\begin{equation}
R_{PC} = R_S + \Gamma(l/a)\frac{\rho}{2a},
\end{equation}

\noindent where $\rho$ is the bulk resistivity and $\Gamma(l/a)$ is a slowly varying function of order unity. Inelastic processes still influence the nonlinear conductance, although multiple elastic scattering events broaden the spectral features.

In the thermal regime $a \gg l$, electrons undergo many scattering events within the constriction. Joule heating dominates transport. The effective local temperature satisfies approximately\cite{duif1989point}

\begin{equation}
T_{\mathrm{eff}}^2 = T_0^2 + \frac{V^2}{4L},
\end{equation}

\noindent where $T_0$ is the bath temperature and $L$ is the Lorenz number. Nonlinearities in this limit reflect temperature dependent resistivity. This regime is useful for identifying phase transitions and critical phenomena.

Therefore, in point contact spectroscopy, the applied voltage fixes the quasiparticle energy scale. Inelastic scattering at that energy modifies the current. By analyzing the nonlinear $I$-$V$ characteristics and their derivatives, one extracts the characteristic energy scales of elementary excitations coupled to electrons. The technique does not measure band dispersion, it measures energy dependent quasiparticle scattering.

Among various applications of point contact spectroscopy, point-contact Andreev reflection spectroscopy (PCAR) has become particularly important for probing  metal-superconductor  interfaces\cite{duif1989point,naidyuk2005electrical,groll2015point,samuely2023point}. It gives important information about the superconducting energy gap and its momentum-space symmetry\cite{gonnelli2008evidence,daghero2011directional}, spin polarization in ferromagnets\cite{shiga2017spin,woods2004analysis,rana2023high}, and exotic states in topological materials\cite{borisov2016high}. The method is based on the phenomenon of Andreev reflection, first proposed by Andreev in 1964\cite{andreev1964thermal}, which governs charge transport across a normal metal-superconductor (N-S) interface. PCAR spectroscopy has been widely employed to study elemental and complex superconductors\cite{monish2025tip}, including conventional and unconventional superconductors\cite{gonnelli2008evidence,Miyoshi2005}, multigap superconductors\cite{szabo2001evidence,tortello2010multigap}, heavy-fermion systems\cite{park2008andreev}, half-metallic ferromagnets\cite{upadhyay1998probing,ji2001determination,nadgorny2001origin},topological superconductor \cite{lee2019perfect},Dirac semimetal \cite{aggarwal2016unconventional,wang2016observation}, and other strongly correlated electron systems. More recently, point-contact spectroscopy helped realization of novel tip-induced superconducting (TISC) phases that appear in a confined region under the point contacts and are confirmed by spectroscopic measurements\cite{lee2019perfect,monish2025tip,monish2025above}.  Thus, point contact spectroscopy is powerful tool for studying various exotic properties of quantum materials.

In many modern quantum materials, the relevant excitation energies lie in the sub millielectronvolt range. To resolve such features, thermal broadening must be minimized so that $k_B T \ll eV$. This requirement motivates measurements in dilution refrigerators, where millikelvin temperatures and high magnetic fields enable high resolution mapping of quasiparticle scattering processes and collective excitations.

In this work, we present a detailed description of the instrumentation and data acquisition procedures for a point-contact spectroscopy setup capable of operating at ultra-low temperatures down to 30~mK using a  dilution refrigerator (DR).
\section{Point contact spectroscopy set up}
\subsection{Fabrication of point contacts}

\begin{figure}[]
    \centering  \includegraphics[width=1\columnwidth,trim=6cm 6cm 6.5cm 4cm,
  clip]{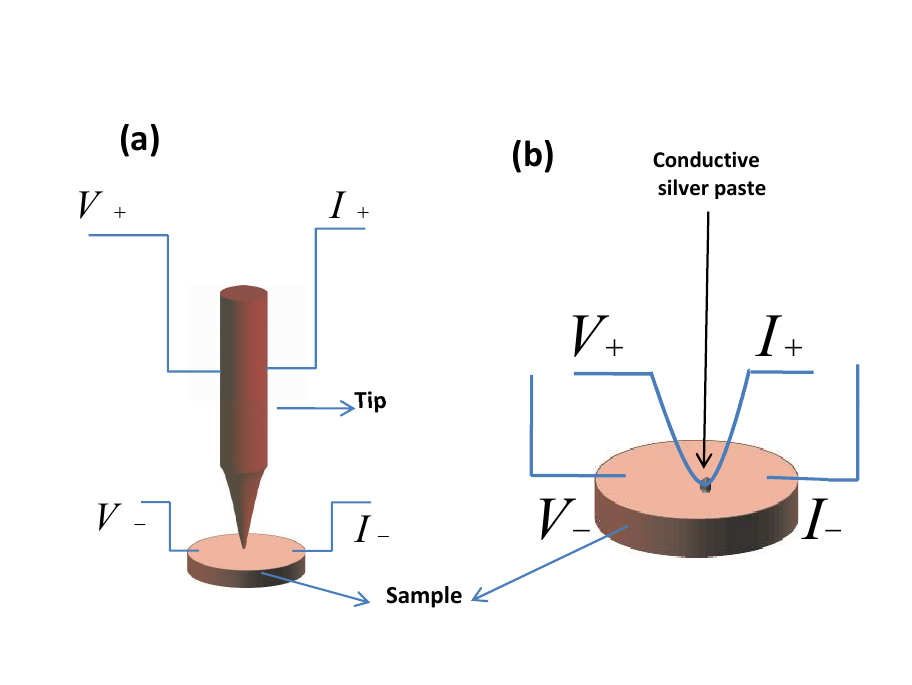}
    \caption{Schematic of point contacts. (a) Needle–anvil technique; (b) Soft point contact.}
    \label{fig:schematics}
\end{figure}

Point-contact spectroscopy relies on the formation of a mesoscopic constriction between a metallic tip and a conducting sample. Several methods have been developed to create such contacts, each with definite advantages depending on the experimental requirements. The most commonly employed techniques are the mechanical needle-anvil method and the soft point-contact technique \cite{naidyuk2005electrical, sheet2004role}, schematically illustrated in Fig.~\ref{fig:schematics}.

In the soft point-contact technique [Fig.~\ref{fig:schematics}(b)], a small droplet of conductive silver paint or paste is mounted between a fine metallic wire (typically gold) and the sample surface. Once it is cured, this droplet forms multiple parallel conductive nano-channels through which electric current can be injected. This gives a stable electrical contact without applying significant physical pressure to the sample. The effective contact diameter and consequently the contact resistance can be modified by applying controlled current or voltage pulses, which either destroy existing conduction paths or create new ones at the junction. While this method is largely non-destructive and provides excellent mechanical stability, it offers limited control over the contact diameter. Achieving a purely ballistic contact regime involves multiple trials and is not always guaranteed. On the other hand, the mechanical needle-anvil technique [Fig.~\ref{fig:schematics}(a)] employs a sharp metallic tip that is brought into direct contact with the sample surface using a precision positioning mechanism. Traditionally, differential screw mechanisms were used for this purpose, but modern experiments, specifically under sub-kelvin temperatures, piezoelecctric nanopositioners are used. Such cryogenic piezo-driven positioners provide highly reproducible motion with step sizes on the order of a few nanometers even at ultra-low temperatures. This helps in a fine and continuous control of the contact resistance thereby enabling controlled transition between different regimes of mesoscopic transport. These advantages make the needle-anvil method very attractive. Other techniques for forming point contacts include lithographically defined point contacts \cite{van1996adjustable}, which also offer precise geometrical control  but require extensive sample preparation. 

Here, we employ the needle–anvil method using a cryogenic piezoelectric nanopositioner. Detailed discussions of the design and implementation of the PCS setup, the PCS electrical measurement , and the data analysis procedure are provided in the following sections.

\subsection{Design and Implementation of a
Needle–Anvil Setup for dilution fridge}
 The PCS setup was  installed in a wet dilution refrigerator (Janis Research, USA). The base temperature of the dilution refrigerator (DR) at the mixing chamber plate is 30 mK, while the sample temperature reaches approximately 35 mK, as measured by a calibrated  ruthenium oxide thermometer mounted near the sample stage. The cryostat is integrated with a cryogenic reliquefier (Cryomagnetics) to reduce liquid helium consumption. Furthermore, the reliquefier and the dilution gas handling system, which are major sources of mechanical vibration, are both electrically and mechanically isolated to achieve a low-noise background.

In the Janis DR system, to minimize the effective thermal load, the cryostat wiring is made resistive and thermally anchored at multiple stages of the DR. This results in a resistance of approximately $R \sim 150~\Omega$  between the electrical feedthrough port at the top of the cryostat and the shuttle. This presents two major challenges:
\begin{enumerate}
    \renewcommand{\labelenumi}{(\alph{enumi})}
    \item The piezoelectric actuator cannot be efficiently driven through such a highly resistive path and would otherwise require prohibitively high voltages to achieve even small displacements, and 
    \item  High-voltage operation in an ultra-low-temperature environment can introduce undesirable heating due to current dissipation along the resistive wiring, thereby disturbing the millikelvin temperature stability.
    
\end{enumerate}
 To overcome these challenges, we adopted special strategies and enabled low-voltage piezo actuation through the given  resistive path. In the following subsections, we will discuss the hardware design and adaptation of a conventional piezo-walker to mitigate the issues mentioned above. 
\subsubsection{The Shuttle}
\begin{figure*}[t!]
    \centering
    \includegraphics[width=\textwidth,trim=0cm 1.3cm 0cm 0cm, clip ]{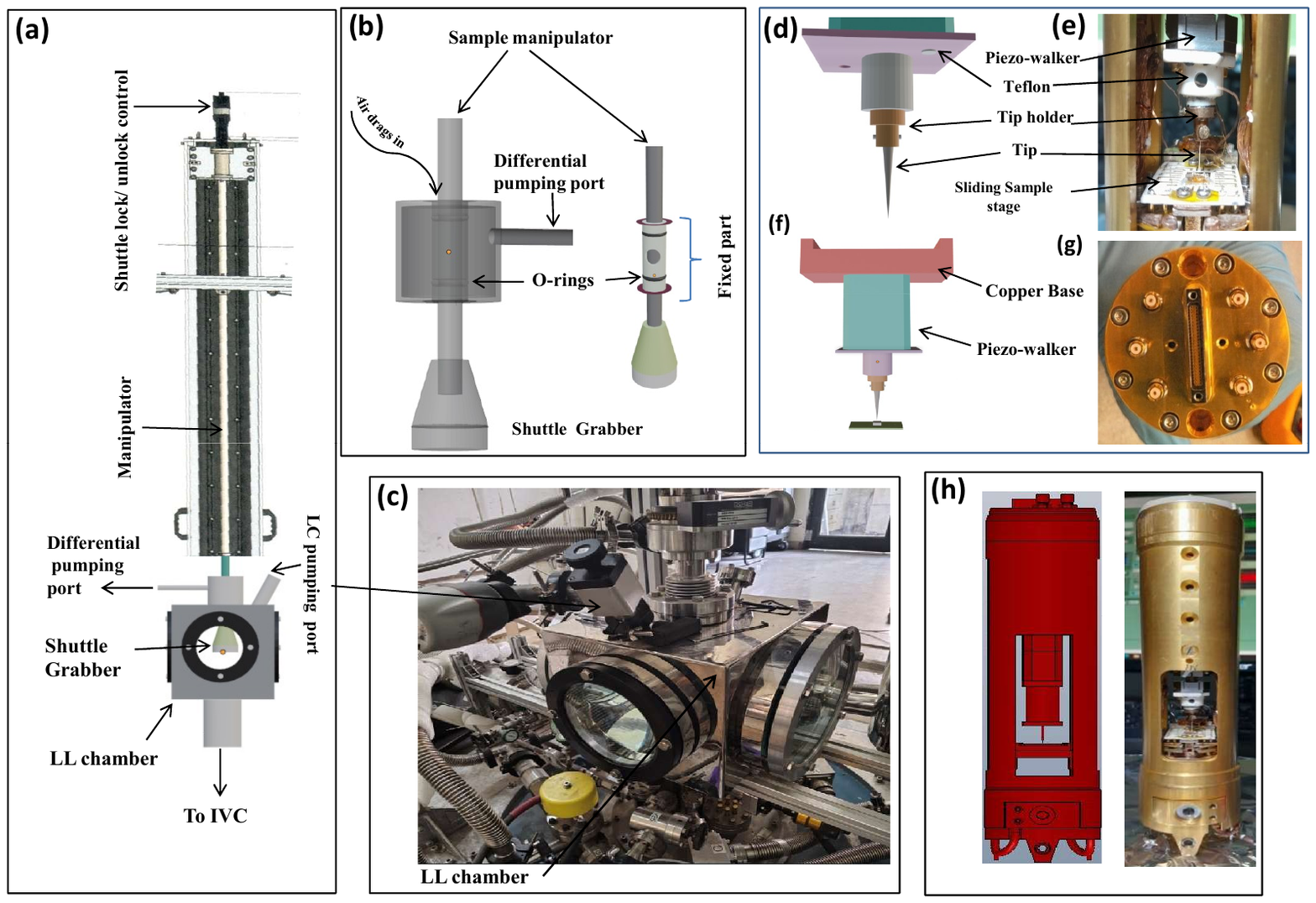}
    \caption{
The shuttle and nano-positioner are  designed for operation at ultra-low-temperature. (a) Schematic of the manipulator with load-lock chamber. (b) Detailed transparent view of the differential pumping section. (c) Real image of the load-lock chamber. (d,e) The  sample–tip assembly mounted on the piezowalker, which is  mounted on an OFHC-grade copper base. (f) Zoomed view of real image of the mounted  piezowalker with the tip–sample assembly inside the  shuttle. (g) View of shuttle micro-D connector. (h) Complete shuttle  (left) and corresponding real image (right).
}
    \label{fig:Probe details}
\end{figure*}

To enable efficient sample and tip exchange without warming the entire cryostat, the system is designed around a top-loading interchangeable shuttle concept. The shuttle, shown in Fig.~\ref{fig:Probe details}(h), is a compact module having an assembly designed for reliable operation down to ~30 mK and under high magnetic fields. The body of the shuttle is machined from oxygen-free high-conductivity (OFHC) copper to ensure excellent thermal conductivity and efficient thermal anchoring to the mixing chamber. It hosts a sample stage consisting of a printed circuit board (PCB), a thermometer, a heater, and a piezo-walker for vertical tip positioning.
The piezo-driven walker  is rigidly mounted on a gold-coated OFHC copper plate as shown in Fig.~\ref{fig:Probe details}(f), to ensure both thermal and mechanical stability. Electrical isolation between the walker and the tip holder is achieved using a 2 mm thick teflon spacer, while the tip holder is fixed to the spacer using Stycast epoxy, as shown in Fig.~\ref{fig:Probe details}(d,e,f). All electrical connections are routed through a micro-D connector at the base of  the shuttle to the room-temperature electronics, as shown in Fig.~\ref{fig:Probe details}(g).

\subsubsection{Adaptation of the piezo-walker to the environment}
The Attocube piezo-walker operates via a "slip–stick" mechanism. The walker is driven by a sawtooth waveform  ($f = 200$~Hz) generated by an Attocube piezo controller (ANC300) . In an ideal scenario (i.e., when there is little to no resistive path between the piezo-walker and the voltage source), the slow ramp of the waveform produces a sticking phase, while the rapid drop generates a strong inertial “kick” that triggers the slip event. However, due to the finite resistance of the wiring, the voltage across the piezo, V(t), is governed by.
\begin{equation}
RC\,\frac{dV(t)}{dt} + V(t) = V_{\mathrm{in}}(t),
\label{eq:rc_final}
\end{equation}
where the piezo behaves as a capacitor ($C \sim 1~\mu$F at room temperature), giving a time constant $\tau \approx 150~\mu$s. To overcome this resistive path and to mitigate heating issues when driving the piezo-walker at high voltages,  the static friction between the walker and the monorail shaft was manually reduced. This adjustment also lowers the dynamic friction, enabling the piezo-walker to operate effectively at lower drive voltages, which is necessary for such ultra-low-temperature environments. Fig.~\ref{fig:Piezoresponse_at_temp}(a) shows a comparison between the applied waveform $V_{\mathrm{in}}(t)$ (red) and the measured voltage at the piezo terminals $V(t)$ (blue) at room temperature (in the regime where the walker is already in motion, i.e., \ after reducing the clamping force or friction). Clearly, a distortion of the waveform is observed. The fast falling edge of the applied signal exhibits a voltage drop of $\Delta V \approx 19.65$~V within $\Delta t \approx 10~\mu$s, corresponding to
\begin{equation}
\left|\frac{dV_{\mathrm{in}}}{dt}\right| \sim 2 \times 10^{6}~\mathrm{V/s},
\end{equation}
whereas at the piezo terminals the same transition is broadened to $\Delta V \approx 16.68$~V over $\Delta t \approx 690~\mu$s, yielding
\begin{equation}
\left|\frac{dV}{dt}\right| \sim 2.4 \times 10^{4}~\mathrm{V/s}.
\end{equation}
This nearly two orders of magnitude reduction demonstrates the strong suppression of high-frequency components due to $RC$ filtering.

\medskip
The displacement of the piezo is given by $x(t) = d_{\mathrm{eff}} V(t)$, and the inertial force responsible for slip is:
\begin{equation}
F_{\mathrm{in}}(t) = m\,d_{\mathrm{eff}}\,\frac{d^2 V(t)}{dt^2}.
\end{equation}
where m is the  mass of the moving stage, including the tip and tip-holder and $d_{\mathrm{eff}}$ stands for the effective piezoelectric coefficient.\\
Slip occurs only when this inertial force exceeds the static friction force $F_s$:
\begin{equation}
m\,d_{\mathrm{eff}}\,\frac{d^2 V(t)}{dt^2} > F_s.
\label{eq:slip_final}
\end{equation}

\medskip
 Before the clamp is loosened or friction is reduced , the static friction $F_s$ is large enough that the above condition is not satisfied at any point during the cycle, resulting in a purely stick regime with no net motion. The data shown in Fig.~\ref{fig:Piezoresponse_at_temp}(a) correspond to the case where the clamping force is reduced such that  Eq.~(\ref{eq:slip_final}) is satisfied over a finite time interval within each cycle.

\medskip
To note, even in the walking regime, the rounded waveform implies that the condition for slip is fulfilled only over a restricted time window $\Delta t_{\mathrm{slip}} \ll T$:
\begin{figure*}[t!]
    \centering
    \includegraphics[width=\textwidth,trim=0cm 1.2cm 0cm 0cm, clip ]{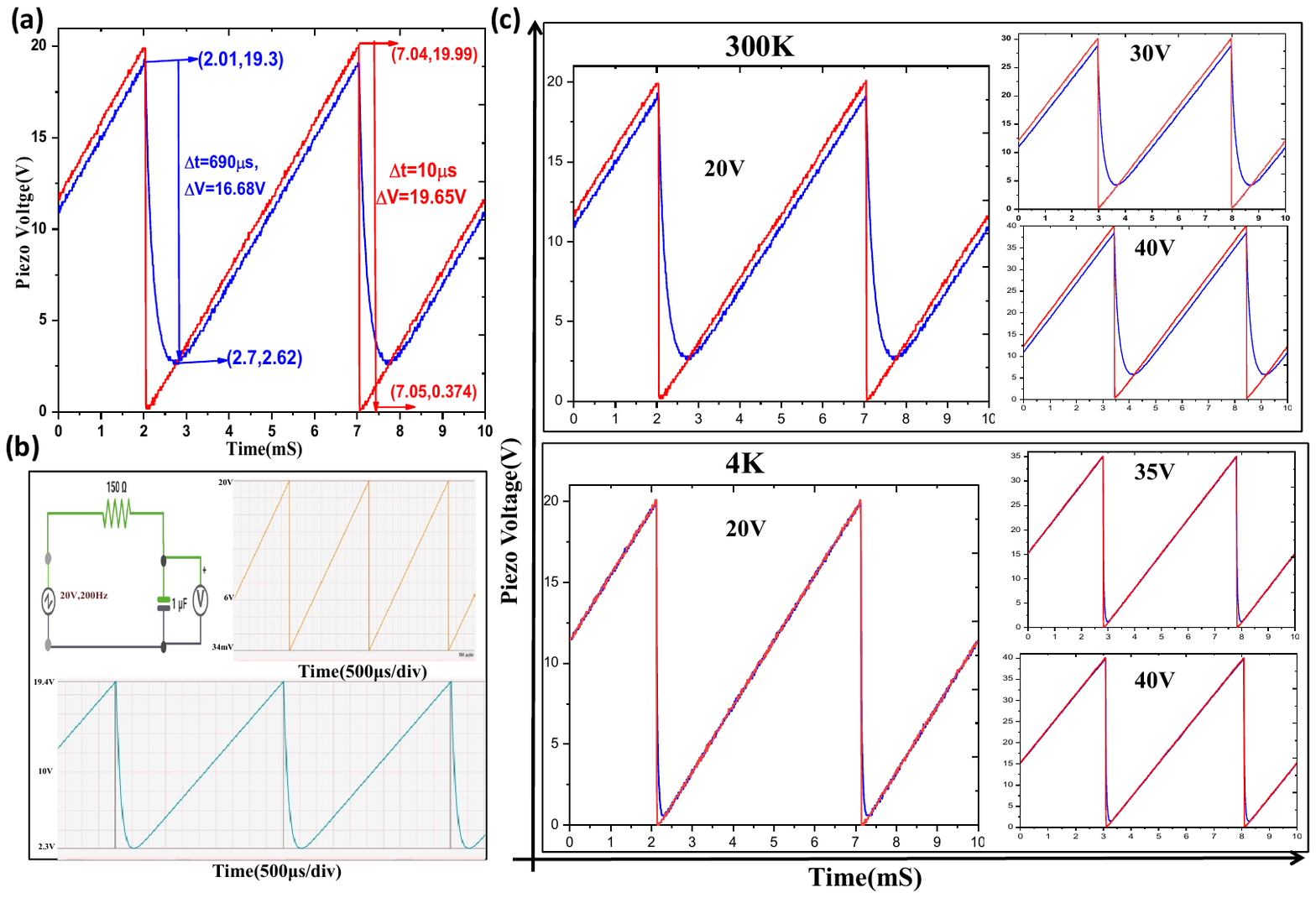}    
      \caption{ (a) Applied 20 V, 200 Hz ,sawtooth waveform (red line) and measured voltage across the piezo terminals (blue line), showing the reduced voltage across the piezo due to wire resistence. (b) Equivalent simulated  circuit diagram and simulated circuit response showing the input sawtooth signal (top-right ) and voltage across the equivalent  capacitor (bottom). (c) Voltage across the piezo-terminal at different sawtooth  voltages (20V, 30V/35V, 40V) at 300 K (top panel) and 4 K ( bottom panel) temperatures.}
    \label{fig:Piezoresponse_at_temp}
\end{figure*}
\begin{equation}
\Delta t_{\mathrm{slip}} = \left\{ t \in [0,T] \;\bigg|\; m\,d_{\mathrm{eff}}\,\frac{d^2 V(t)}{dt^2} > F_s \right\}.
\end{equation}

Once slip is initiated, the motion is governed by the kinetic friction force $F_k$, with $F_k < F_s$. However, reducing the clamping force simultaneously lowers both $F_s$ and $F_k$. As a result, although motion becomes possible, the maximum force that can be sustained by the walker is limited by the static friction\cite{das2024compact},
\begin{equation}
F_{\mathrm{max}} \lesssim F_s.
\end{equation}

\noindent This introduces an inherent trade-off: loosening the clamp facilitates motion under reduced excitation conditions but reduces the load-bearing capability of the walker and limits the maximum force applied by the tip to the sample.

\medskip
The degree of waveform distortion is further influenced by the temperature dependence of the piezoelectric capacitance. As the temperature decreases, the dielectric constant of the piezoelectric material is reduced, leading to a decrease in capacitance and, consequently, a reduction in the $RC$ time constant. This improves the electrical bandwidth of the system.
This effect is experimentally observed in Fig.~\ref{fig:Piezoresponse_at_temp}(c), where the waveform measured at low temperature exhibits significantly sharper edges compared to room temperature, indicating reduced distortion. From Eq.~(\ref{eq:rc_final}), a decrease in capacitance directly enhances both $\frac{dV}{dt}$ and $\frac{d^2 V}{dt^2}$ at the piezo. This increases the inertial force and expands the fraction of the cycle over which the slip condition is satisfied. Thus, in addition to mechanical tuning of friction, the intrinsic reduction of capacitance at low temperature provides a natural mechanism for improving slip--stick efficiency, partially compensating for the limitations imposed by the series resistance of the cryogenic wiring. Therefore, we have successfully adapted the piezo walker to function reliably at ultra-low temperature condition of DR.

\subsubsection{Room-Temperature Shuttle  Transfer and Docking Procedure}
The transfer of the shuttle from room temperature to the mixing chamber is a critical process accomplished by a dedicated top-loading system, which consists of  a vertical manipulator, load-lock (LL) chamber, and a gate valve separating the LL from the DR's inner vacuum chamber (IVC) schematically shown in Fig.~\ref{fig:Probe details}(a,b,c).
The procedure begins by mounting the fully assembled shuttle onto the shuttle grabber at the head of the manipulator. The manipulator is a long, hollow stainless steel tube that provides the rigid mechanical link needed for precise vertical insertion. The LL chamber was  pumped down to a pressure of \( 1\times 10^{-5} \) mbar, matching the high vacuum inside the DR's IVC. Once this vacuum is achieved, the gate valve is opened. The manipulator is then slowly lowered using a pulley system. During this descent, a critical differential pumping mechanism is employed: a roughing pump connected to a port between two Viton O-rings on the manipulator shaft shown in Fig.~\ref{fig:Probe details}(b), removes any minute amount of air that might be dragged down by the moving shaft, ensuring that the  vacuum of the DR space is maintained . Successful docking of the shuttle into its receptacle at the bottom of the mixing chamber is confirmed by monitoring the  resistance of a sensor or jump in mixing chamber temperature.

\begin{figure}[h!]
    \centering
    \includegraphics[width=1\columnwidth,trim=6cm 7cm 7cm 7cm,
  clip]{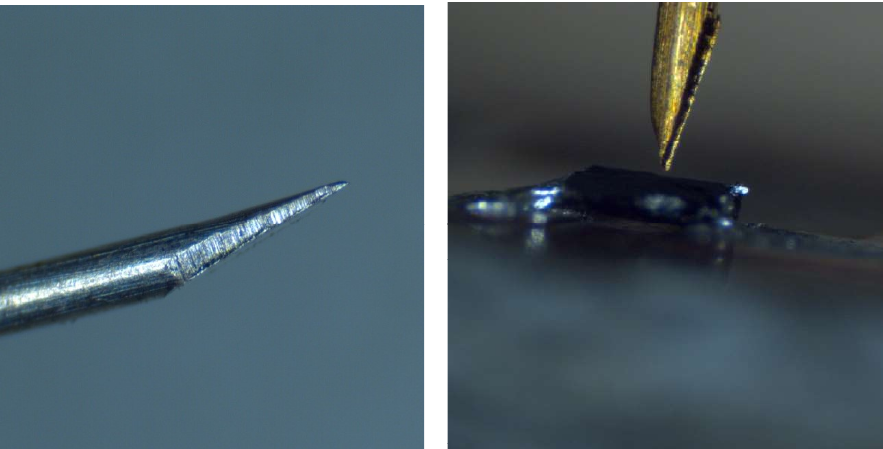}
    \caption{Real image of a 0.25 mm diameter Ag tip (left) used for PCS, and the tip aligned on top of a sample (right).}
    \label{fig:tipimages}
\end{figure}

The sample is mounted on a dedicated sample stage within the shuttle. It is affixed using GE varnish (GE Varnish C5-101, Oxford), which provides excellent electrical insulation while maintaining good thermal conductivity, ensuring that the sample is well thermalized. The sample stage is equipped with a resistive heater and a calibrated ruthenium oxide (RuO) temperature sensor for local temperature control and monitoring. For PCS measurements, metallic tips of 0.25 mm diameter are prepared ex-situ, either by electrochemical etching or by mechanical cutting at an angle, as shown in  Fig.~\ref{fig:tipimages}. 
\subsubsection{Thermalization}
After docking, the manipulator is separated from the shuttle and withdrawn back into the LL chamber, and the gate valve is closed. The insertion of the warm shuttle (at room temperature) inevitably causes a transient temperature rise in the mixing chamber and other stages of the DR. To accelerate re-cooling, a mechanical heat switch is engaged, thermally connecting the mixing chamber stage to the 4 K liquid helium bath. This allows the entire insert, including the shuttle, to equilibrate to approximately 4 K within a few hours. Once this temperature is reached, the heat switch is disengaged, and further cooling to the base temperature of 30 mK is achieved through the condensation  of the \(^3\)He--\(^4\)He mixture. To suppress thermal radiation, two shutters attached to the DR's heat-switch shaft, one located below the IVC flange and another at the still chamber, are kept closed.

\subsection{Electrical Measurements}

\begin{figure}[h]
    \includegraphics[width=1\columnwidth, trim=8cm 5cm 10cm 5cm, clip]{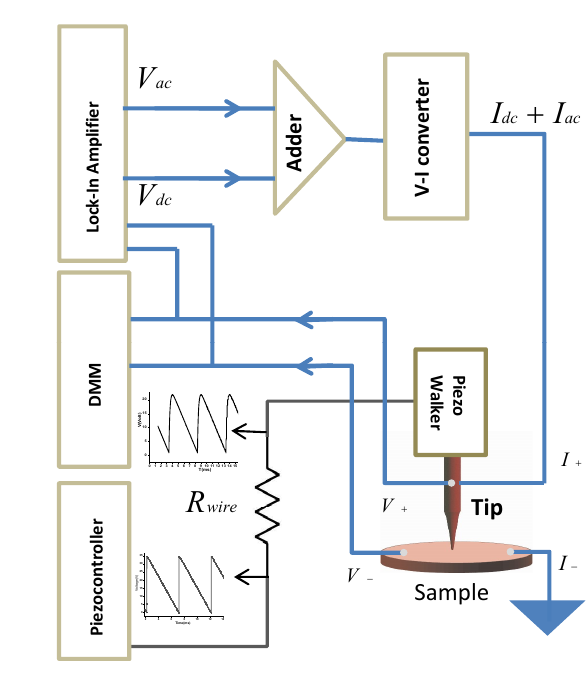}
    \caption{Schematic of the point-contact Andreev reflection (PCAR) measurement setup using a lock-in based modulation technique.}
    \label{fig:PCAR measurement}
\end{figure}

Point-contact Andreev reflection (PCAR) measurements were performed using  a standard AC modulation technique with a lock-in amplifier (SR830 DSP, Stanford Research Systems). The SR830 served a dual role, providing both the AC excitation signal (from its internal oscillator) and the DC bias voltage (via its auxiliary output). An adder circuit and active voltage-to-current (V-I) converter , schematically shown in Fig.~\ref{fig:PCAR measurement}, was used to combine the AC and DC voltages  and convert the summed voltage into a current bias. The modulated current consisted of a small AC current  $I_{\mathrm{ac}}$ superimposed on a DC current $I_{\mathrm{dc}}$, which was then passed through the point contact. The tip-sample contact was established by driving the piezowalker via the Attocube ANC300 piezo-controller, enabling precise and controlled variation of the contact size.
The DC voltage \(V\) developed across the point contact was measured directly by a digital multimeter (DMM, Keithley 2000). Simultaneously, the AC voltage response was measured by the lock-in amplifier. The first harmonic response of the lock-in (\(V^{(1\omega)}\)) is proportional to the differential resistance \(\frac{dV}{dI}\) of the junction. The underlying principle is as follows: using a current bias of the form
\[
I(t)=I_{\mathrm{dc}}+I_{\mathrm{0}}\cos(\omega t),\qquad I_{\mathrm{0}}\ll I_{\mathrm{dc}}
\]
the voltage across the junction, \(V = V(I)\), can be expanded in a Taylor series about \(I_{\mathrm{dc}}\):
\[
\begin{aligned}
V(t) &= V\!\left(I_{\mathrm{dc}} + I_{\mathrm{0}}\cos \omega t\right)\\
&\approx V(I_{\mathrm{dc}})
+ \left.\frac{dV}{dI}\right|_{I_{\mathrm{dc}}} I_{\mathrm{0}}\cos \omega t
+ \frac{1}{2}\left.\frac{d^2V}{dI^2}\right|_{I_{\mathrm{dc}}}
I_{\mathrm{0}}^2 \cos^2 \omega t \\
\end{aligned}
\]
The lock-in amplifier referenced at the fundamental frequency \(\omega\) measures a signal proportional to the first-harmonic coefficient:
\[
V^{(1\omega)} \propto \left.\frac{dV}{dI}\right|_{I_{\mathrm{dc}}}\, I_{\mathrm{ac}}.
\]
Consequently, the differential resistance is given by \(\left.\frac{dV}{dI}\right|_{I_{\mathrm{dc}}} \propto V^{(1\omega)} / I_{\mathrm{ac}}\), and the differential conductance, which is the quantity of primary interest in PCAR spectroscopy, is simply its reciprocal:
\[
\left.\frac{dI}{dV}\right|_{I_{\mathrm{dc}}}=\left(\left.\frac{dV}{dI}\right|_{I_{\mathrm{dc}}}\right)^{-1}.
\]
In the small-modulation limit, where \(I_{\mathrm{ac}}\) is sufficiently small, higher-order terms are negligible, and the first-harmonic signal accurately represents \(dV/dI\). The second-harmonic component at \(2\omega\), proportional to \(d^2V/dI^2\), can be monitored to assess nonlinearity or the effects of finite modulation. In PCAR measurements, the effective energy resolution is limited by both thermal broadening and the modulation amplitude; therefore, the modulation amplitude $I_{\mathrm{ac}}$ is typically chosen such that $I_{\mathrm{ac}} \cdot R_N  \lesssim \frac{k_B T}{e}$, where $R_N$ is the normal-state resistance of 
the junction and $e$ is the electron charge.

\begin{figure}[h!]
    \centering
   \includegraphics[width=1\columnwidth,trim=3.5cm 6cm 3.5cm 5cm,clip]{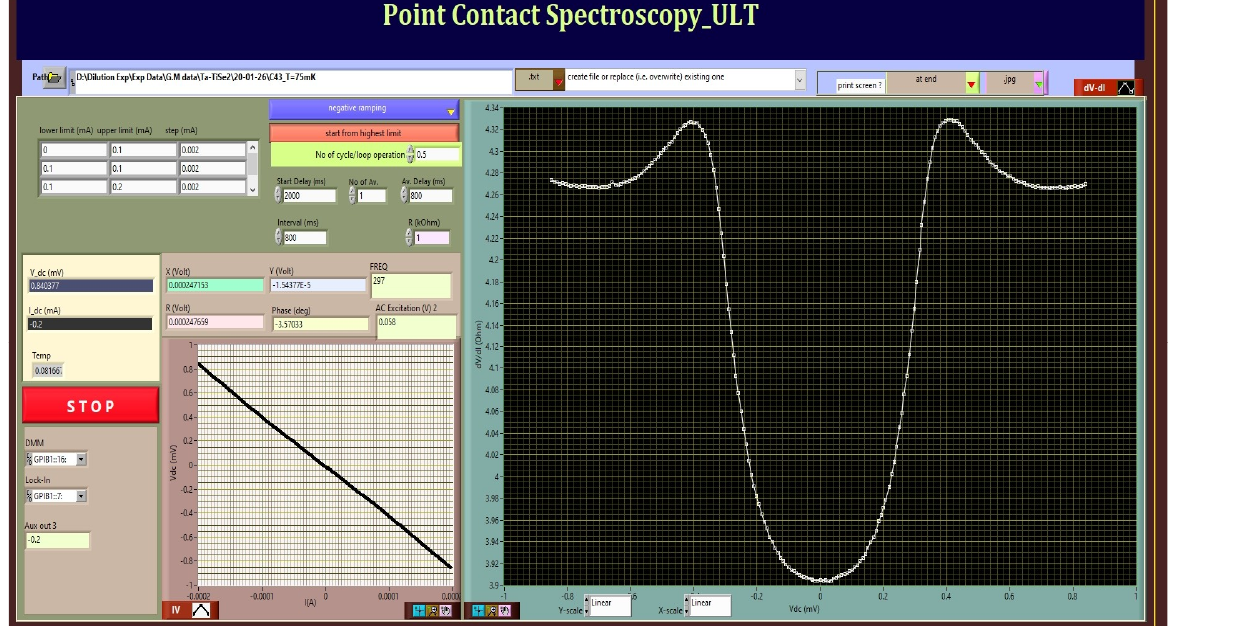}
    \caption{A LabVIEW-based, user-friendly, interactive graphical user interface for automated PCAR measurement and control.}
    \label{fig:labview}
\end{figure}

All data acquisition was automated via a General Purpose Interface Bus (GPIB) and Ethernet connection using a TCP/IP interface, controlled by a custom LabVIEW-based program. The software provides an interactive graphical user interface (GUI), shown in Fig.~\ref{fig:labview}, through which all measurement parameters—including temperature, magnetic field, and bias ranges—can be set and monitored.

\subsection{Data Analysis}

The quantitative interpretation of PCAR measurements is commonly based on the Blonder–Tinkham–Klapwijk (BTK) theory \cite{blonder1982transition}, which describes transport across normal-metal--superconductor (N--S) interfaces by introducing a dimensionless interfacial barrier parameter \(Z\). This parameter accounts for the probability of Andreev reflection versus ordinary reflection at the interface. Extensions of the original BTK model have been developed to incorporate additional physical effects relevant to realistic materials. These include the incorporation of spin polarization \(P\) for ferromagnetic materials \cite{strijkers2001andreev}, a quasiparticle lifetime broadening parameter \(\Gamma\) to account for inelastic scattering and finite quasiparticle lifetimes \cite{dynes1978direct}, and extensions for multiband superconductivity \cite{chen2001evidence}. These models enable a detailed and quantitative analysis of experimental conductance spectra.

To analyze our data, the measured differential conductance (\(dI/dV\)) spectra are first normalized to their high-bias value. Theoretical \(dI/dV\) curves are then generated using the appropriate BTK-based model and fitted to the experimental data. The key fitting parameters include the superconducting energy gap \(\Delta\), the effective junction temperature \(T\) (which accounts for thermal broadening), the barrier strength \(Z\), the quasiparticle broadening parameter \(\Gamma\), and the spin polarization \(P\), where applicable. The fit is optimized to extract these physical parameters, providing insight into the superconducting and electronic properties of the material under study.
\\
\section{Results and Discussion}
Using this setup, the PCAR measurements were performed on a single crystal of tantalum-doped titanium diselenide (Ti$_{1-x}$Ta$_x$Se$_2$, $x = 0.2$) with a silver tip. Titanium diselenide (TiSe$_2$) is a layered transition-metal dichalcogenide that exhibits a well-known charge density wave (CDW) transition near $\sim200$\,K \cite{di1976electronic}. In its pristine form, TiSe$_2$ is  a semimetal or a small band-gap semiconductor, and extensive studies have focused on understanding the origin of the CDW state and its relationship with excitonic and electron–phonon interactions\cite{rossnagel2011origin}. When electron doping is introduced, such as through elemental substitution, intercalation, or pressure, the CDW order can be weakened or suppressed, often giving rise to new electronic phases including superconductivity\cite{morosan2006superconductivity,kusmartseva2009pressure}.

\begin{figure}[h!]
    \centering
    \includegraphics[width=0.9\columnwidth,trim=0cm 0.7cm 0cm 1cm,clip]{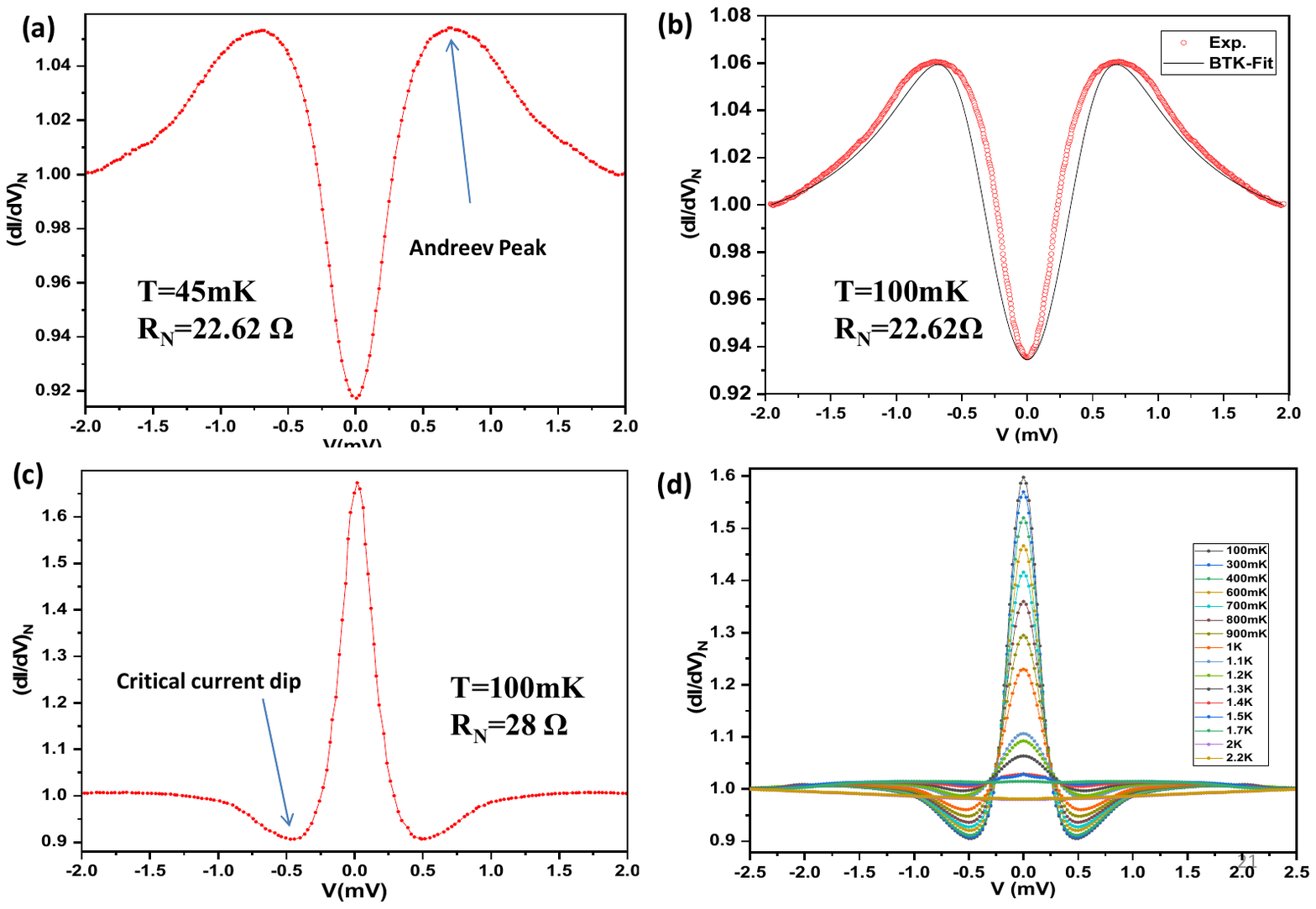}
    \caption{PCAR spectra of Ag/Ta-TiSe$_2$ .(a) Normalized  conductance spectra obtained in the ballistic regime.(b) Normalized conductance spectra (red circles) and the corresponding  BTK fit (black line). (B) Normalized thermal conductance spectra. (d) Temperature dependence of the thermal spectrum.}
    \label{fig:spectrum}
\end{figure}

Here, we present PCAR measurements on Ta-doped TiSe$_2$, synthesized elsewhere~\cite{manna2025quasi}, with a superconducting transition temperature $T_c = 2.2$\,K. Figure~\ref{fig:spectrum}(a) shows the  PCAR conductance spectra obtained in the ballistic regime showing Andreev peaks. (b) Conductance spectra fitted with the BTK model (black line). Figure~\ref{fig:spectrum}(c) shows the spectrum in the thermal regime, and its temperature dependence is presented in Fig.~\ref{fig:spectrum}(d). The temperature dependence of the thermal spectra shows that the superconducting spectral features gradually diminish with increasing temperature and disappear completely at $T = 2.2$\,K, consistent with the bulk $T_c$ of the material.
All the spectra are normalized with respect to the conductance at higher bias, i.e.,
\[
\left(\frac{dI}{dV}\right)_N = \frac{dI/dV}{\left.(dI/dV)\right|_{V \gg \Delta/e}},
\]
where $\Delta$ is the superconducting energy gap.

\section{Conclusion}

We have successfully integrated a point-contact spectroscopy (PCS) setup with a top-loading dilution refrigerator, enabling stable measurements at ultra-low temperature. The system employs a needle-anvil geometry driven by a cryogenic piezo-nanopositioner, which allows \textit{in-situ} formation of mesoscopic point contacts with precise control over the contact resistance. To overcome the limitations imposed by high-resistance cryostat wiring, we implemented a strategy to reduce mechanical friction in the slip-stick mechanism. These measures permit reliable piezo operation at reduced voltages, ensuring minimal heating and stable millikelvin temperatures.

The performance of the setup was validated by PCAR measurements on a Ta-doped single crystal \(\mathrm{TiSe_2}\) (\(T_c \approx 2.3~\mathrm{K}\)). The obtained conductance spectra clearly resolve the superconducting gap, and the spectral features systematically diminish with increasing temperature consistent with the bulk superconducting transition. This demonstrates the capability of the instrument to perform high-resolution quasiparticle scattering spectroscopy in the millikelvin regime.

\begin{acknowledgments}
We thank Dr. Ravi Prakash Singh, IISER Bhopal, for providing the single crystals of Ti$_{1-x}$Ta$_x$Se$_2$. G.M. thanks the University Grants Commission (UGC), Government of India, for the Senior Research Fellowship (SRF). A.M. thanks IISER Mohali  for the Senior Research Fellowship (SRF).We also thank Anmol Anand  for his involvement in the early
stages of this work. G.S. acknowledges financial assistance from the Science and Engineering Research Board (SERB), Govt. of India (grant number: CRG/2021/006395). The metallic body of the shuttle and manipulator was designed in collaboration with Janis Research Company, USA. The authors declare no competing financial interests.
\end{acknowledgments}
\newpage
\bibliographystyle{apsrev4-1}
\bibliography{Instru.bib}
\end{document}